
\def\IR{{\hbox{{\rm I}\kern-.2em\hbox{\rm R}}}}
\def\IB{{\hbox{{\rm I}\kern-.2em\hbox{\rm B}}}}
\def\IN{{\hbox{{\rm I}\kern-.2em\hbox{\rm N}}}}
\def\IC{{\ \hbox{{\rm I}\kern-.6em\hbox{\bf C}}}}

\def\IZ{{\hbox{{\rm Z}\kern-.4em\hbox{\rm Z}}}}

\def\underarrow#1{\vbox{\ialign{##\crcr$\hfil\displaystyle
{#1}\hfil$\crcr\noalign{\kern1pt
\nointerlineskip}$\longrightarrow$\crcr}}}
%

\input phyzzx
\sequentialequations
\overfullrule=0pt
\def\part{\not\hskip-3pt\partial}
\rightline{IASSNS-HEP-92/24}
\vglue-.50in
\font\bigbold=cmbx12 scaled\magstep2
\title{\bigbold TWO DIMENSIONAL STRING THEORY
\break  AND BLACK HOLES\foot{Lecture at the conference on
``Topics in Quantum Gravity,'' University of Cincinnati, April 3-4, 1992.}}
\vglue-.25in
\author{Edward Witten}\foot{Research supported by NSF Grant No.
91-06210.}
\address{Institute for Advanced Study
\break Olden Lane
\break Princeton, NJ 08540}
\vglue-.25in
\fourteenpoint
\abstract{This lecture surveys a few loosely related topics,
ranging from the scarcity of quantum field theories -- and the role that
this has played, and still plays, in physics
-- to paradoxes involving black holes
in soluble two dimensional string theory and the question of whether naked
singularities might be of even greater interest to string theorists
than black holes.}
\endpage

Consistent quantum field theories are apparently scarce.  As far as we
understand, truly consistent four dimensional field theories (without
Landau ghosts) require Yang-Mills gauge fields, so as to ensure
asymptotic freedom.  This assertion is an extrapolation beyond what we
firmly know, but is supported in various simple examples by mathematical
theorems and computer experiments, mostly about $\phi^4$ theories.

The scarcity of quantum field theories is arguably one of the most
important things we teach our students, and one of the main assets
physicists have had in understanding elementary particles.  In fact, it
has been very fortunate for the progress of physics that a generic
Lagrangian, such as a Fermi-type Lagrangian
$$
{\cal L} = \int d^4 x[\overline \psi i\hskip-4pt\part \psi
- G_F ~\overline \psi
\gamma_\mu \psi~ \overline \psi \gamma^\mu \psi]
\eqn\joe
$$
does not (presumably) lead to a consistent quantum theory.  The search
for a consistent quantum theory that reduces to the Fermi theory at low
energies led to the discovery of the now-standard electroweak theory,
and with it an appreciation of the role of non-abelian gauge fields in
physics.  The scarcity of quantum field theories also accelerated the
discovery of QCD.  Of course, questions of consistency also were crucial
in the discovery of Maxwell's equations and of special and then general
relativity.

General relativity -- which is based on a geometric concept somewhat
analogous to that of non-abelian gauge theory -- is the one real
experimental signal we have of physics at energies way beyond
accelerator energies.  Other experiments sensitive to physics
of ultra-high energies -- notably proton decay and neutrino mass
measurements, and searches for magnetic monopoles and other
ultra-massive big bang relics -- have so far given null results, albeit
important ones.

The peculiar properties of general relativity are of course intimately
tied up with the fact that Newton's constant has dimensions of
(Mass)$^{-2}$.  This leads among other things to the perturbative
unrenormalizability  of the theory
$$
{\cal L} = M^2_{P\ell} \int d^4x \sqrt g ~ R
\eqn\jim
$$
which I will take at face value as an indication that the quantum theory
does not exist.  I consider this the optimistic as
well as most likely interpretation.

Indeed, the fact that general relativity is unrenormalizable is the one
reason that we have some hope of learning something about new physical
principles that prevail at higher-than-accelerator energies.  If,
conversely, this problem were solved -- say in the strongest form of
showing that at a nonperturbative level an arbitrary Lagrangian coupling
general relativity to matter can be successfully quantized -- we would
feel truly forsaken by what in the twentieth century has become our best
friend, the goddess of consistency.  Our hopes of learning what the
right theory is, or more modestly of learning about fundamental new
structures in physics, would greatly dwindle.

Taking then at face value the appearance that general relativity is
unrenormalizable, let us ask what new consistent framework could have
conventional quantum theory, and Riemannian geometry, as limiting cases.
 Since geometrical structures of the relevant depth do not exist just by
accident, this kind of question cannot be expected to have many possible
answers -- whether right or wrong for physics.  In fact, no insight has
ever been gained by frontal assault.  Happily, a possible answer -- in
the form of string theory -- emerged as an unintended consequence of
work on (unsuccessful models of) strong interactions.

What we really understand about string theory are rules for
perturbative computations of scattering amplitudes.  Curiously
enough, these rules are much simpler, in key ways, than standard Feynman
rules.  It can be argued that the tree level scattering of gauge bosons
and gravitons -- first worked out by Feynman, De Witt, and others in
the 1960's -- is most easily calculated by embedding
Yang-Mills theory or gravity
in a string theory, that is, by calculating appropriate string
amplitudes and then taking the appropriate limit to extract the field
theory result.  The advantage of embedding the conventional calculations
in string theory is all the greater at the one loop level.  This became
clear in the calculations by Green and Schwarz of graviton scattering in
the early 1980's, and has recently begun to be exploited systematically
in work of Bern, Kosower, and others.  Indeed, the simplification
obtained by embedding the conventional calculations in string theory is
substantial enough that this method may well become standard in
computations of QCD processes relevant to the SSC.  It may also lead to
new insights about QCD.  In any event, the simplification that arises
this way in the standard calculations in the most interesting
geometrical theories is one of the symptoms of the power of the
mysterious new geometrical structure that underlies string theory.

What is missing in our knowledge of string theory is precisely that this
structure is mysterious -- that we are far from understanding the
conceptual-geometrical, and presumably Lagrangian, framework from
which all this is
to be derived.  Roughly, we know the Feynman rules, but we do not
understand the classical theory from which they came.   Nor are we
close, as far as I know.  But the spinoffs that have been discovered,
even in purely geometrical problems (index theory on loop spaces, soluble
conformal field theory and knot invariants, mirror symmetry, cohomology
of moduli space of Riemann surfaces -- just to name a few) show the
exceptional richness of this structure.

In groping toward an answer, we grapple with the following paradigm.
The string theory analog of a metric is a two dimensional Lagrangian
$$
{\cal L}(X) = {1 \over 8\pi} \int_\Sigma d^2\sigma \sqrt h ~
\left[h^{\alpha\beta} \partial_\alpha X^\mu~\partial_\beta X^\nu g_{\mu\nu}
(X^\lambda) + \ldots \right]~.
\eqn\jack
$$
(The $X$'s are a map from a two dimensional surface $\Sigma$ to a space-time
manifold $M$ of arbitrary dimension.)

The string theory analog of a solution of the Einstein equations is then
an ${\cal L}(X)$ for which the corresponding quantum theory is {\it
conformally invariant}.  Then on the space of ${\cal L} (X)$'s $ - \ie$, the
space of two dimensional quantum field theories -- one should find a
function (perhaps some version of the $c$ function of Zamolodchikov) that
is stationary precisely when ${\cal L}(X)$ defines a conformal field
theory.

This paradigm is quite beautiful.  It is perhaps vaguely reminiscent of
an early idea of Penrose that the metric structure of spacetime should
be coded in properties of the space of null geodesics.  Here null
geodesics -- one dimensional objects -- are replaced by two dimensional
minimal surfaces in spacetime (or more generally, by surfaces that are
stationary points of ${\cal L}(X)$).

In one direction, the paradigm is quite correct and extremely useful.
Every conformal field theory (obeying a couple of simple conditions)
determines a classical solution of string theory.  But the paradigm has
trouble in the reverse direction, because it clashes with the
fundamental first lesson that I recalled at the outset.  There is no
problem in writing down a generic ${\cal L}(X)$, but the generic ${\cal
L}(X)$ does not determine a quantum theory; there is therefore no known
way to define the beta function associated with a generic ${\cal L}(X)$,
or to discuss the vanishing of this beta function.  The problem is
particularly acute if one tries to think about sigma models with time
dependence (leading to world-sheet operators of negative dimension) or
superficially unrenormalizable operators  (corresponding to the
characteristic massive modes of the string).

This clash between what string theorists appear to need and the rest of
what we know in physics is in my opinion the main obstacle to progress
in string theory;  I cannot emphasize this enough. It is a hint that
two dimensional field theory as we know it is not an adequate framework.
I have long suspected that we need something cruder, that would contain
less information in return for less work.  This suspicion was one of my
main motivations, incidentally, in developing topological  field theory.
 Today, however, I will be sketching some other developments.

Though we
do not yet know the end of the story, a main focus of effort by many
physicists in the last few years has been to experiment with ${\cal
L}(X)$'s in a simple situation -- that in which {\it space-time}, as
well as the world-sheet, is two dimensional.  The simplicity that arises
here, though very unexpected technically, is perhaps crudely analogous
to the simplicity that appears in many types of problems in two
dimensional mathematical physics.

My discussion of these matters will be purely qualitative.  For more
detail I refer to some of the excellent review articles, or to the
generally well-known original papers (including the celebrated papers of
Brezin and Kazakov, Douglas and Shenker, and Gross and Migdal on the
double scaling limit for $D < 2$).

Granted that a generic ${\cal L}(X)$ does not lead to a quantum field
theory, what are some of the good ${\cal L}(X)$'s that we know with a
two dimensional target space?  The simplest is free field theory, with a
linear coupling added to achieve conformal invariance:
$$
{\cal L}(X^0, X^1) = {1 \over 8\pi} \int d^2\sigma \big[-\partial_\alpha X^0
\partial^\alpha X^0
+\partial_\alpha X^1 \partial^\alpha X^1
- 2 \sqrt2 R^{(2)} X^1 \big]~.
\eqn\jake
$$
The linear term is a little strange.  It means that in this two
dimensional world, it is impossible to achieve Poincar\' e invariance. We
will offer an intuitive explanation of this later.  I will call this
puzzle $(a)$.  Several other puzzles will appear presently.

A slightly more sophisticated example, still leading to a conformal
field theory, is obtained by adding an experimental interaction, the
so-called Liouville term:
$$
\widetilde{\cal L}(X) = {\cal L}(X) + \mu \int d^2\sigma ~e^{-\sqrt2 X^1}~.
\eqn\joss
$$
For $\mu > 0$, this repels us from the region of $X^1 \rightarrow -
\infty$ where, technically, the string coupling is large.

Once one finds a classical solution -- whether in general relativity or
string theory -- one would like to understand the scattering theory
around this classical solution.  In general this is difficult.  The
greatest progess in $D = 2$ string theory has come from the remarkable
discovery that scattering theory around the particular classical
solution of string theory determined by $\widetilde{\cal L}(X)$ is exactly
soluble.  It can be described in terms of a degenerate gas of free
fermions interacting with an upside-down harmonic oscillator potential:
$$
V(\lambda) = - {1 \over 2} \lambda^2~.
\eqn\jessie
$$
This was discovered by studying the quantum mechanics of an $N \times N$
matrix, on the one hand by diagonalizing the matrix and constructing the
Hamiltonian, and on the other hand by expanding in Feynman diagrams.

The relation of $\widetilde {\cal L}(X)$ to the free fermions leads to
several additional puzzles about this theory (beyond puzzle $(a)$ that I
have already noted above):

$(b)$ The physics of $\widetilde {\cal L}$ is really related just to the
behavior of the free fermions for $\lambda \rightarrow \infty $.  What
does the second region at $\lambda \rightarrow - \infty$ have to do with
it?

$(c)$  The free fermions are an (elementary) integrable system. They have
infinitely many conserved qualities.  If one considers a state with
incoming fermions of energies (relative to the fermionic surface)
$\varepsilon_1, \ldots, \varepsilon_k$ then
$$
Q_n = \sum_{i=1}^k ~\varepsilon_i^n
\eqn\jason
$$
is conserved for each $n$.  Of course, $Q_1$ is just the Hamiltonian.

$(d)$  More generally, the canonical transformations of the free fermion
phase space enter in deeper study of the model.

By now, $(c)$ and $(d)$ have been pretty well explained from a two
dimensional field theory point of view, by studying certain discrete
operators of the two dimensional theory.  What about $(a)$ and
$(b)$?  To
explain these points, it is judicious to remember that the interest of
string theory derives from its interpretation as a theory of space-time
gravity.  Look for a more general Lagrangian
$$
\hat{\cal L}(X^0, X^1) = {1 \over 8\pi} \int d^2\sigma ~
\sqrt h ~\big[h^{\alpha\beta}
\partial_\alpha X^\mu~\partial_\beta X^\nu g_{\mu\nu}(X^\lambda)
+ \Phi(X^\lambda) R^{(2)}\big]~.
\eqn\jock$$
In a one loop approximation, it is easy enough to solve the equations of
conformal invariance.  One finds that the target space metric
$g_{\mu\nu}$ must (up to a coordinate transformation) take the form
$$
ds^2 = {du~ dv \over 1 - uv}
\eqn\john
$$
that is characteristic of a black hole.  Indeed \john\ is the essence of
the Schwarz-schild metric (dropping the angular variables and an
inessential conformal factor) in a form that exhibits its maximal
analytic continuation.  The two asymptotically flat regions of the black
hole are the two regions of $uv$ large and negative, the exterior of the
light cone in figure (1).

This apparently explains our problem $(b)$ about the free fermions of the
matrix model.  The unexpected region $\lambda \rightarrow - \infty$ of
the free fermions apparently corresponds to the second asymptotically
flat world that is discovered upon making a maximal analytic
continuation of the original Schwarzschild solution, which as we recall
is

$$
ds^2 = - (1 - {2GM \over r}) dt^2 +  \big(1 -
{2GM \over r} \big)^{-1}dr^2 + r^2 d \Omega^2 ~.
\eqn\jontue
$$

What about problem $(a)$, the lack of Poincar\' e invariance in our two
dimensional target space?  Actually, in writing the metric in equation
\john, I have not indicated the
dilaton field, and so I have not exhibited the one parameter of the
solution, which is the mass $M$ of the black hole, and is buried in the
possibility of adding a constant to the
dilaton field.  (In the brand of two dimensional target space gravity
that is determined by this model, the dilaton field enters in the ADM
mass formula.)  One may ask what happens to the solution as $M
\rightarrow 0$.

A four dimensional Schwarzschild black hole goes over to Minkowski space
for $M \rightarrow 0$; but a charged black hole in four dimensions
cannot have $M < Q$, and goes over for $M \rightarrow Q$ to a
non-trivial and interesting limit, the extreme Reissner-Nordstrom black
hole.  I believe that the lack of Poincar\' e invariance in two
dimensional string theory can be understood intuitively by making an
analogy with the $s$-wave sector of a Reissner-Nordstrom black hole in
four dimensions.  It is as if there is a charge sitting at $X^1 = -
\infty$; the linear dilaton field in \jake\ is the field of this
charge, and the simple solution \jake\ is the analog not
of Minkowski space
but of an extreme Reissner-Nordstrom black hole.

Now actually one can do better and find an exact black hole solution in
this two dimensional world, by using $SL(2,\IR)$ current algebra.  To
put it differently, and I think this is a vivid illustration of how rich
and physical string theory is, the black hole solution pops out of a
simple, universal calculation using current algebra and gauge theory.

In the black hole analogy, the Liouville term $\mu \exp({-\sqrt 2 X^1})$
in $\widetilde
{\cal L}$ corresponds to a repulsive non-gravitational interaction that
does not have a really good counterpart in four dimensions.

So far I have outlined, or at least alluded to, answers to our four
questions $(a)$-$(d)$.  But at this point,
two more basic questions arise:

(1)  Is the black hole that arises here stable, and if so how is this
compatible with Hawking's discovery that conventional black holes
radiate?

(2)  If standard two dimensional string theory is -- as I believe -- a
theory of matter interacting with the analog of an extreme
Reissner-Nordstrom black hole, why is this not more obvious in studies
of the free fermion model?

I will propose answers.  I must warn you, however, that the answer to
(2) will be frustrating, though in my opinion entirely consistent and
even correct.

For (1), I want to first step back and ask, can a black hole be in
equilibrium with matter, and can such an equilibrium state be described
by a pure quantum mechanical state?   My answer to this is not really novel,
and involves
considerations of Hartle and Hawking from the 1970's -- with
a twist at
the end that depends on the linear dilaton field of the two dimensional
string theory.  (My comments might also be compared to observations by
Seiberg and Shenker in a recent Rutgers preprint.)

To begin with, we ask whether one can have pure quantum mechanics of any
kind in the field of a black hole.  There is no problem here, as long as
one considers both asymptotically flat ends.  One picks a Cauchy
hypersurface $S$, as sketched in figure (1).  In the standard way, one
defines a quantum Hilbert space ${\cal H}$ by quantization on this
hypersurface.  It is convenient to pick $S$ to be a hypersurface of time
reflection symmetry, as sketched in the figure.

What would be a natural state vector in ${\cal H}$?  In general, in a
time dependent situation such as the field of the black hole, such a
question has no answer; there is a natural Hilbert space, but because of
particle creation and annihilation, there is no distinguished ``vacuum.''
The black hole is special, however.  Recall the Euclidean black hole
solution.  In two dimensions it is described by a metric of the form
$$
ds^2 = dr^2 + \tanh^2 r~ d\phi^2
\eqn\jerry
$$
and looks like a semi-infinite cigar.  In four dimensions, there is a
similar formula, with additional angular variables that we suppress.
The Euclidean black hole has a hypersurface $\widetilde S$ of time symmetry,
given by $\phi = 0, \pi$.  $\widetilde S$ has the same extrinsic and
intrinsic geometry as $S$, and so we could equally well regard $S$ as
the boundary not of half of the extended Lorentzian Schwarzschild space
but of a half of its Euclidean analog, as sketched in figure (2).  We
will use the letter $W$ to denote the relevant half of the Euclidean
black hole metric, with boundary $S \cong \widetilde S$.

We can now try to define a distinguished state vector $\psi \in
{\cal H}$ by following a recipe similar to that of Hartle and Hawking.
We denote the quantum field variables on $S$ as $X$ and those on $W$ as
$Y$.  We attempt to define a vector $\psi(X) \in {\cal H}$ by
integrating over the $Y's$:
$$
\psi(X) = \int_{Y|{_S}=X} DY~ \exp\left( - {\cal L}(Y)\right)~.
\eqn\jan
$$
We will postpone temporarily the discussion of whether the integral is
well-behaved, and consider first the physical properties of $\psi$ if it
is.

The key point is that although the two asymptotically flat ends, say
$S_L$ and $S_R$, of $S$ are asymptotically infinitely distant in the
Lorentzian black hole, in the Euclidean hole the distance between them
is finite.  In fact they are separated asymptotically by a distance
$\beta/2$, with $\beta$ being the asymptotic circumference of the cigar.
 Thus if $X_L$ and $X_R$ are quantum field variables on $X_L$ and $X_R$,
the wave function $\psi$ is approximately (if one considers only
observables supported in the asymptotically flat region)
$$
\psi(X_L, X_R) = \langle X_L | \exp(- {1 \over 2} \beta H) | X_R
\rangle ~,
\eqn\joel
$$
with $H$ the Hamiltonian.

This is a pure state, but if one wants to consider only observables
supported on one end, say $S_R$, then one must integrate out $X_L$ to
form a density matrix in $X_R$.  In the approximation of $\joel$, the
density matrix in question is simply
$$
\eqalign{
\rho (X_R{}^\prime, X_R) &= \int DX_L ~\overline \psi(X_R{}^\prime,
X_L)~ \psi(X_L, X_R)\cr
&= \langle X_R{}^\prime | \exp(-\beta H)| X_R\rangle ~.\cr}
\eqn\jaffe
$$
This is a thermal state at the Hawking temperature $T = 1/\beta$.

Thus, though $\psi$ is a pure state, it looks like a mixed, thermal
state to an observer at one end.  To such an observer, this state
appears to describe a black hole in thermal equilibrium with matter, the
outgoing Hawking radiation being in balance with the incoming flux of
thermal radiation.

Is such equilibrium possible?  As we will see, the answer is no in the
conventional four dimensional world, but yes in two dimensional string
theory.   The key point is simply to ask whether the integral \jan\ converges.
In any dimension, the one loop approximation to this integral is
well-defined.  In that approximation, one finds a thermal energy density at
temperature $T = 1/\beta$.  In infinite volume, the total energy due to
this thermal energy density is infinite.

In the two loop approximation, we will begin to see the gravitational
back-reaction of the infinite thermal energy.  In four dimensions, this
will produce gravitational collapse on a larger scale, showing that
thermal equilibrium between a black hole and matter is impossible in
four dimensions.  (The same conclusion is sometimes reached by
considering the negative specific heat of a black hole in four
dimensions.)

In two dimensional string theory, there is a very elementary but crucial
difference.  This comes from the linear dilaton term in \jake. As a
result of this term, the gravitational coupling vanishes exponentially
for $X^1 \rightarrow \infty$.  As a result of this, the gravitational
effects of the infinite thermal gas are
finite!   So the perturbative
contributions to
the path integral \jan\ will all be convergent, and there is no
difficulty presumably in defining the state $\psi$.  Thus, in two
dimensional string theory, it is possible to have a pure quantum state
which to an observer at one end appears to describe thermal equilibrium
between a black hole and matter.

Not only is this possible; it is presumably what one gets by
systematically calculating string loop corrections to the $SL(2,\IR)/
U(1)$ conformal field theory of the black hole.  In fact, it is
presumably what one would get in any calculation of Lorentzian black
hole physics which can be obtained by analytic continuation of a
Euclidean black hole calculation.  This would be so for the standard
conformal field theory, in which the Lorentzian and Euclidean black
holes are described by $SL(2,\IR)$ cosets that are related by analytic
continuation.  There may well be an analog of the free fermion
description of the black hole interacting with matter.

Now as a preliminary to addressing question (2), let me discuss a crude
version of that question that is often  asked.  Can ``the'' black hole,
that is the $SL(2,\IR)$ coset model, decay to the standard two
dimensional string model, with world-sheet Lagrangian ${\cal L}$ or
$\widetilde{\cal L}$? (I put quotes around the word ``the,'' since I have
suggested above that the standard ${\cal L}$ or $\widetilde{\cal L}$
describes ``a'' black hole-like object.) The answer is clearly no.
``The'' black hole does not decay, since it is in equilibrium with the
thermal gas around it.

Here is another naive preliminary to question (2).  Can ``the'' black
hole be created in the usual scattering theory around the
standard ${\cal L}$
or $\widetilde{\cal L}$ states?  The answer again is obviously no.  In usual
scattering theory, one excites the ground state by a finite energy
disturbance, but ``the'' black hole lies above the ground state (the
${\cal L}$ or $\widetilde{\cal L}$ vacuum) by an infinite energy.
Therefore, it cannot be created in conventional scattering theory.

This infinite energy may seem worrisome, so let me make a further
comment.  In two dimensional string theory, the spacetime gravitational
and dilaton fields approach their flat space values for $X^1 \rightarrow
\infty$ exponentially fast.  The analog of the ADM mass is the
coefficient of $\exp(-X^1)$.  The infinite energy that was crucial above
merely means that the coefficient of $\exp(-X^1)$ grows linearly in $X^1$
for $X^1 \rightarrow \infty$, and thus the ominous-sounding infinite
thermal energy just means that the correction to the asymptotic vacuum
behavior is proportional to $X^1 \exp(-X^1)$ instead of $\exp(-X^1)$.  The
extra factor of $X^1$ obviously does not change things drastically.

Having disposed of some preliminaries, let us now discuss what I regard
as the most incisive version of question (2): why have not black hole
effects, such as Hawking radiation, been noticed in scattering theory
around the ground state?  For I have suggested that this scattering
theory involves interaction of matter with a black hole analog.

I will give an answer that I consider convincing but frustrating -- it
will show the price we pay for embedding black hole physics in a soluble
model.

We noted above that the description by free fermions leads to
conservation of infinitely many charges
$$
Q_n = \sum_{i=1}^k ~\varepsilon_i^n~.
\eqn\jonesy
$$
These charges are all coupled to gauge fields (as one sees by examining
the discrete excitations of ${\cal L}$), so the $Q_n$ values of a black
hole are measurable outside the horizon and so well-defined.

Thus there must be a family of black hole solutions depending on
infinitely many parameters $Q_n$; ``the'' black hole
(described by $SL(2, \IR$) current algebra) is just a special case.  The
more general black holes would be related to more general ${\cal
L}(X)$'s containing superficially unrenormalizable operators, so they
are hard to study.

In the theory of Hawking radiation, there is a classical potential
conjugate to every conserved charge carried by the black hole.
(Electric and magnetic charge are the examples most often considered in
four dimensions.)   A black hole carrying non-zero values of these other
classical potentials does not just radiate thermally.  How it radiates
depends on the values of the chemical potentials.

Now consider scattering theory around the standard $D = 2$ string
vacuum.  The initial state is a black hole-like object with some value
of the $Q_n$'s, presumably zero.  We excite it by sending in some
particles from infinity.  The values of the $Q_n$'s and hence the
chemical potentials of the system so created depend on what we send in.
Therefore, in a generic experiment, the radiation that will come back
out will not be thermal -- it depends on the chemical potentials of the
``hole'' and hence on what was sent in.   Therefore, generically, one
will not get thermal radiation as a result of exciting the hole by
incoming matter.  This answers question (2).

Is there, however, any way to excite the hole in such a way that it will
emit thermally?  The answer is yes, as I will now explain -- but in a
way that is sure to be frustrating.

If one sends in on the hole a thermal distribution of particles, then
conservation of the $Q_n$'s ensures that a thermal distribution will
come back out.  Therefore, it must be that sending in a thermal
distribution of particles is a way to excite the vacuum to just the
values of the chemical potentials that will lead to thermal radiation.
This is an unglamorous way to ``see'' Hawking radiation, but it seems to
be all one can hope for when the black hole is embedded in an integrable
system.

I wish to add the following remarks.  (1) In a scattering experiment
with just a few incident particles, instead of a macroscopic number,
most probably the resulting chemical potentials have such extreme
values that a thermodynamic description is not valid, even allowing for
the chemical potentials.  (For four dimensional black holes, the
analogous issues have been worked out by Preskill, P. Schwarz, Shapere,
Trivedi, and Wilczek.)
(2)  The Liouville interaction, which I have
not built into these remarks explicitly, is a repulsive interaction that
will scatter many low energy incident particles non-gravitationally.
This is very important quantitatively. (3) Ellis, Mavromatos, and
Nanopoulos have related the $Q_n$'s to quantum coherence of black holes.

Finally, then, I do believe that the {\it standard}  two dimensional
string theory describes the unitary quantum mechanics of matter
interacting with an
object similar to a black hole.  Integrability means that the behavior
is rather different from black hole physics as we usually know it.  The
linear dilaton field means that stable quantum mechanics of the black
hole in thermal equilibrium is possible; it is related to $SL(2,\IR)$
current algebra, and an analog of the free fermion description may well
exist.

In this lecture, I have been focussing on black holes to elucidate
the surprising role of a black hole look-alike in soluble
two dimensional string theory.  There are indeed many unsolved
conceptual mysteries in black hole physics, and string theory may give
a fruitful vantage point for rethinking them.  We should be alert,
however, to other possibly related problems that might be even more
pertinent for string theorists.  Here I have in mind the whole question
of the role in physics of general relativistic singularities other
than black holes.
The ``cosmic censorship'' conjecture of Penrose asserts roughly
(in its original form) that black holes are the only type of singularities
that can evolve in classical general relativity from good initial data.
Attractive though the cosmic censorship
hypothesis is, the evidence for it is quite limited.  We are pretty
well convinced that gravitational collapse that is sufficiently close to
being spherically symmetric leads to black holes; but we know very little
about highly aspherical collapse.
Though a proof
of cosmic censorship would extend the scope of classical general relativity
in a dramatic fashion, its failure would possibly benefit physics
even more, since the breakdown of classical general relativity at a naked
singularity might give us the chance to observe effects of quantum
gravity -- or string theory.

Many physicists seem to think that naked singularities would be ``objects,''
in the same sense that black holes are objects.  I think that we should
be on the lookout for naked singularities as ``events,'' analogous
to instantons rather than to solitons, and perhaps looking to a distant
observer much like a miniature big bang.
In this context, it is very interesting that there are events known
to astrophysicists -- the cosmic gamma ray bursts, whose extragalactic
origin has been pretty well established by the recent BATSE observations --
among whose most {\it conservative} explanations
scenarios involving highly aspherical gravitational collapse
(inspiraling neutron star pairs, for instance) are prominent.  These
events are therefore fairly good candidates as already observed events
in which cosmic censorship may have been violated if it is in fact wrong.
These events ought to give us a good incentive for thinking about what
string theory would have to say if cosmic censorship is false in general
relativity.

\end